\documentclass[article]{revtex4}
\usepackage[dvips]{graphicx,color}
\usepackage{epsfig,epstopdf,ifpdf}
\usepackage{latexsym,slashed,subfigure}
\usepackage{amsmath,amssymb,amsfonts}
\newcommand{\beq}{\begin{equation}}
\newcommand{\eeq}{\end{equation}}
\newcommand{\beqr}{\begin{eqnarray}}
\newcommand{\eeqr}{\end{eqnarray}}
\def\lsim{\raise0.3ex\hbox{$\;<$\kern-0.75em\raise-1.1ex\hbox{$\sim\;$}}}
\def\gsim{\raise0.3ex\hbox{$\;>$\kern-0.75em\raise-1.1ex\hbox{$\sim\;$}}}
\begin{document}

\title{Analytical Derivation of Three Dimensional Vorticity Function for wave breaking in Surf Zone}

\author{R. Dutta \footnote{dutta.22@osu.edu}}
\address{Department of Mathematics, The Ohio State University} 

\maketitle
\medskip

\section{abstract}
In this report, Mathematical model for generalized nonlinear three dimensional wave breaking equations 
was developed analytically using fully nonlinear extended Boussinesq equations to encompass rotational 
dynamics in wave breaking zone.
 The three dimensional equations for vorticity distributions are developed from Reynold based stress equations.
Vorticity transport equations are also developed for wave breaking zone. 
 This equations are basic model tools for numerical simulation of surf zone to explain wave breaking phenomena. 
The model reproduces most of the dynamics in the surf zone. Non linearity for wave height predictions is also shown
close to the breaking both in shoaling as well as surf zone. 

{\bf Keyword} Wave breaking, Boussinesq equation, shallow water, surf zone. 
PACS : 47.32-y. 

\section{Introduction}
Wave breaking is one of the most complex phenomena that occurs in the
near shore region. During propagation of wave from deep to shallow
water, the wave field is transformed due to shoaling. Close to the
shoreline, they become unstable and break. In the process of breaking,
energy is redistributed from fairly organized wave motion to small
scale turbulence, large scale currents and waves. \\
Classical Boussinesq theory provides a set of evolution equations for sueface water waves 
in the combined limit of weak nonlinearity (characterized by $\delta \le 1 $) and weak
dispersion ($\mu \le 1$) with the raio $\cfrac{\delta}{\mu^{\rm 2}}=O(1)$. the parameters represent a 
wave height to water deth ratio and a water depth to wavelength ratio, respectively. \\
It has been shown by numerous researchers that Boussinesq-type
equations for varying water depth can describe nonlinear
transformation in the shoaling region quite well. In the last couple
of decades, a lot of research effort has gone into improving the
predictive capability of these equations in the intermediate
water-depth and close to the surf zone (see e.g. Nwogu \cite{nwogu} ,
Madsen \cite{madsen}, Wei \cite{wei}). It was established that to extend the
validity of these equations to the deep water, higher order dispersive
terms will have to be retained. To improve the predictive capability close to wave breaking,
 the higher order nonlinear terms are very important to include in the equation.
However, all these models
use additional terms that is artificially added to the momentum equation,
which would then reproduce the main characteristic of a breaking wave,
i.e. the reduction in wave height to describe wave breaking phenomena.
Wave breaking in the software FUNWAVE (FUNWAVE is based on the model described by Nwogu \cite{nwogu})
is modeled by introducing momentum mixing term developed by Kennedy et
al \cite{kennedy}. \\
Starting with the
works of Nwogu \cite{nwogu} and Madsen \cite{madsen}, most progress have been 
done to explain wave breaking phenomena. Shen \cite{shen} addresses the problem with partially rotational flow.
Vorticity dynamics and formation into the fluid is very important in the wave breaking as well 
surf zone. To address this problem, Veeramony \& Svendsen \cite{veeramony2}
derived breaking terms in Boussinesq equation assuming flow as a two-dimensional rotational
flow. Here, the breaking process is modeled by assuming that vorticity is generated in the roller 
region of the breaking wave and solving vorticity transport equation to determine the distribution of the
vorticity. This naturally introduces additional terms in the momentum
equation which causes wave height reduction as well as changes in the
velocity field. However, since this model is based on stream function formulation, it cannot be
trivially extended to three-dimensional flow. The phenomena of wave
breaking in Boussinesq equations are being modeled using quite few
techniques which can preserve the wave shape as well as include energy
dissipation mechanism.  Shen \cite{shen} developed a generalized form
of Bousinesq equation in three dimensional by introducing  vertical flow field with arbitrary
vorticity distribution up to O($\mu^{\rm 2})$. But he did not describe
momentum transport equation with full description of rotational flow.
Recently, Zou et al \cite{zou} addressed the problem by including the
higher order terms in Boussinesq equation for two dimensinal flow. This model
solves vorticity distribution based on the parametric form taken
form surface roller experimental data.  In this paper, we developed a general analytic
form for breaking term for fully nonlinear set of Boussinesq equations
for three dimensional vertical flow field near surf zone region.
Derivation of breaking term from Reynold stress based vorticity
transport equation was also developed to describe rotational field as
a complete model for thorough understanding of wave breaking phenomena. \\ The term nonlinear indicates
that no truncation based on powers of $\delta$ is employed. 
The paper is organized as follows: Section 2 discusses the
basic governing equations for continuity and momentum with
boundary conditions. Section 3 describes the equation for horizontal
and vertical velocity distribution for potential and rotational
components.  In section 4, the breaking term is derived for velocity
transport equation for fully nonlinear case and solved analytic form vorticity
transport equation from Fourier series expansion.
Reynold stresses are analyzed in the breaking region. In last
section, results were discussed with conclusion. A parametric analysis of the role eddy viscosity
profile is being discussed in detail.

\section  {Basic Equations}
We consider a three-dimensional wave field with free surface
$\eta(x,y,t)$ propagating over a variable water depth $h(x,y)$.  As we
are primarily concerned with wave breaking, we only consider here wave
propagation in shallow water. Wave in this region can be characterized
by two non-dimensional parameters $\delta = {a}/{h}$ and $\mu =
{h}/{l} $ where $a$ is the characteristic wave amplitude and $l$ the
characteristic wave length.  The parameter $\mu$ is a measure of
frequency dispersion and $\delta$ that of the non linearity of the
wave. In this study, since we are only considering shallow water
waves, we only have to consider weakly dispersive waves (up to $
O(\mu^{ 2})$) but have to retain all nonlinear terms.
In this paper, the variables are non-dimensionalized using following
scaling:
\begin{equation}\label{xx}
\begin{split}
 x= \hat{x}/l,  y = \hat{y}/l, z = \hat{z}/h, \; t = \hat{t}\sqrt{gh}/l, \\ 
\hat{u}= \left( \delta\sqrt{gh} \right)u,  \hat{v} = \left( \delta\sqrt{gh}\right)v, \hat{w} = \left( \delta
\mu \sqrt{gh} \right) w
\end{split}
\end{equation}
where the $\;\hat{}\; $ represents the dimensional variables, $g$ is the
acceleration due to gravity, $u$ and $v$ are the horizontal components
of the velocity in the $x$ and $y$ directions respectively, $w$ is the
vertical velocity. We start with the Eulerian equations of continuity
and momentum in  nondimensionalized form for velocity field $ {\bf u} = (u,v,w)$ as:
\begin{equation}
\begin{split}
\cfrac{\partial u} {\partial t} + {\delta} u\cfrac{\partial u}{\partial
  x} + {\delta}v\cfrac{\partial u}{\partial y} + {\delta}
w \cfrac{\partial u}{\partial z} + \cfrac{\partial p}{\partial x} = 0 \\
\cfrac{\partial v} {\partial t} + {\delta}v \cfrac{\partial v}{\partial
x} + {\delta} v\cfrac{\partial v}{\partial y} + {\delta}
w \cfrac{\partial v}{\partial z} + \cfrac{\partial p}{\partial y} = 0 \\
{\delta \mu^{ 2}} \cfrac{\partial w} {\partial t} + {\delta^{ 2}
\mu^{ 2}} u \cfrac{\partial w}{\partial x} + {\delta^{ 2}\mu^{
2}} v \cfrac{\partial w}{\partial y} + {\delta^{ 2} \mu^{ 2}}
w \cfrac{\partial w}{\partial z} + {\delta} \cfrac{\partial p}{\partial z} + 1 = 0
\end{split}
\end{equation}

Since the fluid flow is rotational, we also have three dimensional
 vorticity field ${\bf s} = (s_x,s_y,s_z)$ in the fluid defined as
\begin{equation}
 {\bf \bigtriangledown} \times {\bf u} ={\bf s}  
\end{equation} where $ \bigtriangledown = (\partial/\partial x, \partial/\partial y,
\partial/\partial z). $ 
The continuity equation then becomes, 
\begin{equation}
\bigtriangledown \cdot u + \cfrac{\partial w}{\partial z} = 0 
\end{equation}
Here $ \bigtriangledown \cdot u = (\partial u/\partial x,\partial v/\partial y). $
The above equations satisfy two boundary conditions for velocity at
bottom and at free surface.  At the free surface $ {\it z}=
\eta(x,y,t)$, since particles are free to move with fluid velocity,
the kinematic boundary condition is
\begin{equation}
w_{\rm \eta} = {\bf u_{\rm \eta}} \cdot {\bf\bigtriangledown}\eta  + \cfrac{\partial \eta}{\partial t} 
\end{equation}and at bottom ${\it z= -h(x,y)} $  
\begin{equation}
 w_b = -u_b \cdot {\bigtriangledown} h   
\end{equation}
where $ {\bf u_{\rm \eta}}=(u_{\rm \eta},v_{\rm \eta}) $ is two component horizontal surface velocity.
 $ \bigtriangledown \eta =(\eta_{\rm x},\eta_{\rm y}), $ 
 $ \bigtriangledown h =( h_{\rm x}, h_{\rm y} )$ refer to horizontal derivative with respect to 
x and y in all subsequent calculations.   The horizontal 
component for vorticity field
$ \bf s =( s_{\rm y}, -s_{\rm x})$ can be described as,
\begin{equation}
\cfrac{\partial u}{\partial z} - \mu^{\rm 2} {\bigtriangledown} w ={\bf s}
\end{equation}
with  ${\bf u} = (u,v)$ as two component horizontal field whereas
vertical component of vorticity expressed as
\begin{equation}
-s_{\rm z} = \cfrac{\partial u}{\partial y} - \cfrac{\partial v}{\partial x}
\end{equation}
This is straightforward calculation from  equation (6) and (8)
which is the beginning equation in three dimensional vorticity 
field formulation.  
\begin{equation}
\mu^{ 2} {\bigtriangledown}^{ 2}w + \frac{\partial^{ 2}w}
{\partial z^{ 2}} = - {\bigtriangledown} \cdot s = S_{ w}
\end{equation}
$w$ represents the vertical velocity of the flow.  
In the above equation, once $w$ solved, horizontal component of velocity
$  u,v $ can be solved from vorticity relation.
In weakly hydro static case ( $ 0 < \mu^{ 2} \ll 1 $ ), solution is
typically obtained from iterative perturbation procedure with
successive correction term up to $\mu^{ 2}$.
 
In case of breaking waves where vorticity is very strong, so $ ({\partial
u}/{\partial z} \sim O(1)). $ We assume solution as, $ u = u_{\rm o}
+ \mu^{\rm 2} u_{\rm 1} + O(\mu^{\rm 4}) $ and $ w = w_{\rm 0} +
\mu^{\rm 2} w_{\rm 1} + O(\mu^{\rm 4}) $ for horizontal and vertical
velocity component.
Under this assumption, Poisson equation becomes 
\begin{equation}
\cfrac{\partial^{\rm 2} w_{\rm 0}}{\partial z^{\rm 2}} = S_{\rm w}
\end{equation}
\begin{equation}
\cfrac{\partial^{\rm 2}w_{\rm 1}} { \partial z^{\rm 2} } = - \left[ 
\cfrac{\partial^{\rm 2}w_{\rm 0}} { \partial x^{\rm 2} } +
\cfrac{\partial^{\rm 2}w_{\rm 0}} { \partial y^{\rm 2} }  \right]
\end{equation}
$ w_{ 0}$,  $ w_{ 1}$ can be calculated from bottom boundary conditions 
using equation (7) separately where the boundary conditions are,  
\begin{equation}
\begin{split}   
 w_{\rm {b0}} = - u_{\rm {b0}} \cdot  {\bigtriangledown}h \\
  w_{\rm {b1}} +  u_{\rm {b1}}\cdot
 {\bigtriangledown}h = 0 
\end{split}
\end{equation} at bottom boundary {\it z = -h }. \\
Since at any other depth $ {\bf z} = z_{\rm r}$, w is constrained by
continuity equation only, so the equation follows
\begin{equation}
\cfrac{\partial w}{\partial z} {\rbrack_{ z}}_{ r}
 = - {\bf \bigtriangledown} \cdot u_{\rm m}  {\rbrack_{\rm z}}_{
 r} + \cfrac{\partial u}{\partial z} \cdot {\bf \bigtriangledown}
 z_{\rm r} {\rbrack_{\rm z}}_{ r} 
\end{equation}
where $u_{\rm m} $ is velocity at any arbitrary depth $z_{ r}$ .
In Boussinesq equation, one may take depth average or any
intermediate velocity for horizontal velocity between bottom and free
surface as reference velocity.  In the wave breaking zone where the
vorticity is developed non uniformly, the equations become simpler
with the choice of depth average velocity which includes contribution
from surface vorticity gradient.  We assume solution for velocity
comes also from rotational contribution due to vorticity at the wave
surface. So the velocity has both potential as well as rotational
component, $ u = u_{\rm p} + u_{\rm r} $ , $ w = w_{\rm p} + w_{\rm r} $ \\
We solve $w_{\rm 0}, w_{\rm 1} $ and $ u_{\rm 0}, u_{\rm 1} $ at any depth $z_{\rm r}$
\begin{equation}
 \frac {\partial w_{ 0}} { \partial z } 
 { ]_{z}}_{ r} = - {\bf \bigtriangledown} \cdot (u_{m}
  - z_{r}  s {\rbrack_{ z}}_{ r}) +
z_{ r}   {\bf \bigtriangledown} \cdot s
\end{equation}  
\begin{equation}
\lbrack  \frac {\partial w_{ 1}} { \partial z } {\rbrack_{ z}}_{ r}=
\lbrack {\bf \bigtriangledown} w_{ 0} {\rbrack_{ z}}_{ r} 
\cdot {\bf \bigtriangledown} z_{ r} 
\end{equation} 
\begin{equation}
\begin{split}
 \cfrac {\partial u_{ 0} } {\partial z} = s
\cfrac{\partial u_{ 1}} {\partial z} =  {\bigtriangledown} w_{ 0} 
\end{split}
\end{equation}
 with boundary condition 
 $ \lbrack u_{\rm 0} {\rbrack_{\rm z}}_{\rm r} = u_{\rm r} $ and $
\lbrack u_{\rm 1} {\rbrack_{\rm z}}_{\rm r} = 0 $ \\
Equations (4) - (16)
form basic shallow water Boussinesq equations.

\section{Equation for horizontal velocity}
In the surf zone, vorticity grows very strongly as a non uniform
function over depth.  Following Shen [2000], we define reference
velocity as $ \tilde u = \bar u + \bigtriangleup \bar u - 
\eta s_{\rm \eta} $ in terms of depth average velocity $ {\it \bar u} $ and
magnitude of vorticity at free surface $ s_{\rm \eta} $ with the
assumption of $ \bigtriangledown \cdot s \ne 0 $.  we set here 
$ z_{\rm r} = \eta $ as linear calibration for    $ z_{\rm r} 
= r(\eta + h) - h $ 
does not hold
here in presence of nonuniform velocity as wave dispersion properties
change both spatially and temporally with vorticity. And boundary
condition can be set as
\begin{equation}
\cfrac{\partial w}{\partial z}|_{\rm \eta} = {\bf \bigtriangledown} \cdot \tilde u
+ \eta({\bf \bigtriangledown} \cdot s_{\rm \eta})
\end{equation}
Integrating equation (9) from bottom to surface and applying boundary
condition to (16) we get $w_{ 0}$ as ,
\begin{equation}
w_{\rm 0} = w_{\rm {b0}} - ( -  {\bf \bigtriangledown} \cdot 
\tilde u + \eta  {\bf \bigtriangledown} \cdot s_{\rm {\eta}} ) H_{\rm z} -
S_{\rm {w0}} 
\end{equation}
where 
$$  S_{\rm {w0}} = \int \int (-\bigtriangledown \cdot s )  dz  dz $$  
is the vertical velocity distribution generated by horizontal
divergence of vorticity added to surface velocity.
Now, once $w_{\rm 0}$ is calculated, $u_{\rm 1}$ can be calculated
from eqn (15) with surface boundary condition $ \{ { u_{\rm 0}
\} }_{\rm \eta} = u_{\rm m} $ and 
$ \{ u_{\rm 1} \}_{\rm \eta} = 0 $
Finally, we calculate horizontal velocity as 
\
\begin{equation}
\begin{split}
u(z)= u_{\rm \eta} - \int_{\rm z}^{\rm \eta} s dz +
{\mu}^{\rm 2} ( S_{\rm {wl}} - \bar S_{\rm {wl}}) \\
+ \cfrac{\mu^{\rm 2}}{2}({H_{\rm \eta}^{\rm 2} } - {H_{\rm z}}^{\rm 2} )  
{\bigtriangledown}(\bigtriangledown \cdot \tilde u - {\eta}{\bigtriangledown} 
\cdot s_{\rm \eta})  \\
+ {\mu}^{\rm 2} ( H_{\rm \eta} - H_{\rm  z}) \left [ 
{\bf \bigtriangledown} ((\tilde u + \eta s_{\rm \eta})\cdot {\bf \bigtriangledown }h )
+ 
({\bf \bigtriangledown} \cdot \tilde u - \eta \bigtriangledown \cdot s_{\rm \eta}) \bigtriangledown h \right ] 
+ O(\mu^{\rm 4})
\end{split}
\end{equation}
which on averaging over depth yields, 
\begin{equation}
\begin{split}
\bar{u} = u_{\rm \eta} - \bigtriangleup \bar{u}  + \cfrac{\mu^{\rm 2}}{3} {H_{\rm \eta}}^{\rm 2} \bigtriangledown
 \left [ \bigtriangledown \cdot \tilde u - {\eta}{\bigtriangledown} \cdot s_{\rm \eta} \right ] 
-  \cfrac{\mu^{\rm 2}}{2} H_{\rm \eta} 
[ {\bigtriangledown}  \tilde{u} + {\eta} s_{\rm \eta} ] \cdot{\bigtriangledown}h \\
- ({\bigtriangledown} \cdot \tilde{u} - {\eta}{\bigtriangledown} \cdot s_{\rm \eta}){\bigtriangledown} h 
+ O(\mu^{\rm 4})
\end{split}
\end{equation} 
$ \bigtriangleup \bar u = \cfrac{1}{H_{ \eta}}
 {\int}_{\rm {-h}}^{\rm \eta} \bigtriangleup u(z) dz $ is the average surface velocity
contribution due to vorticity and it is significant for suspended
sediment particles in the flow.  The term $ \bigtriangleup {\bf u(z)}
= \int_{\rm z}^{ \eta} s dz $ is the change due to depth variation
of vorticity $ {\bf s} $.  The total water depth $ H_{\rm z} $ and
surface elevation $ H_{ \eta} $ are taken as $H_{\rm z}$ = z + h
and $H_{\rm \eta} = \eta + h $.
 The contribution for velocity has 
and rotational component apart from potential due to vorticity 
generation. \\
After we redefine  $  H_{\rm \eta}=d $ and $ z={H_{ z}}/{H_{\rm \eta}} $, 
we express potential and rotational component up to order $ O(\mu^{\rm 2}) $ as,

\begin{equation}
\begin{split}
u_{ p}(z) =\bar u_{ p} + \cfrac{\mu^{ 2}}{2} (\cfrac{1}{3} - z^{ 2}){d^{ 2}} 
 \bigtriangledown( \bigtriangledown \cdot \bar u_{ p}) \\
+ {\mu^{ 2}} (\cfrac{1}{2} - z)d  \left[ \bigtriangledown(\tilde u_{ p} \cdot 
\bigtriangledown h) + ( \bigtriangledown \cdot \tilde u_{ p}) \bigtriangledown h \right] \\
u_{ r}(z) =  \tilde u_{ r} - \bigtriangleup u(z)  + \eta s_{ \eta} + \mu^{ 2} 
( S_{ {wl}}  -  \bar S_{ {wl}}) \\
- \cfrac{\mu^{ 2}}{2} (\frac{1}{3} - z^{ 2})d^{ 2} {\eta} \bigtriangledown
(\bigtriangledown \cdot s_{ \eta})  \\
- {\mu^{ 2}} (\cfrac{1}{2} - z)d {\eta} \left [
(\bigtriangledown \cdot s_{ \eta})\bigtriangledown h -
\bigtriangledown(s_{ \eta} \cdot \bigtriangledown h) \right ]
\end{split}
\end{equation}

Similar expressions for vertical velocity are  
\begin{equation}
\begin{split}
w_{ p}(z) =- (h+z) \bigtriangledown \cdot u_{\rm  p}(z) \\  
= -  \bigtriangledown \cdot \left [ (h+z) \tilde u_{\rm  p} \right ]
- \cfrac{\mu^{ 2}}{2} ( (\cfrac{1}{3}-z^{ 2}) \bigtriangledown \cdot \left [
 d^{ 2} (h + z) (\bigtriangledown (\bigtriangledown 
\cdot \tilde u_{ p}) \right ] \\
-  \mu^{ 2}(\frac{1}{2}-z) \left [ \bigtriangledown \cdot (h+z) ( 
\bigtriangledown (u_{ p}
\cdot \bigtriangledown h) + (\bigtriangledown \cdot u_{ p}) 
\bigtriangledown h) \right ]
\end{split}
\end{equation}

\begin{equation}
\begin{split}
w_{ r}(z) =-\bigtriangledown \cdot (h+z) \tilde u_{ r} -
 \bigtriangledown \cdot \left [ (h+z) \eta s_{ \eta} \right ] \\
-  \cfrac{\mu^{ 2}}{2} \left [  \bigtriangledown \cdot (h+z) 
(\cfrac{1}{3} - z^{ 2})d^{ 2}
{\bigtriangledown}(\eta \bigtriangledown \cdot s_{ \eta}) \right ] \\
+ {\mu^{ 2}} \bigtriangledown \cdot \lbrack ( \frac{1}{2} -z)d
\left [ \bigtriangledown(\eta s_{ \eta}.\bigtriangledown h) 
- ( \bigtriangledown \cdot \eta s_{ \eta})\bigtriangledown h \right ]
\end{split}
\end{equation}

\section{ Breaking Model [Our case: fully nonlinear] } 
The most obvious approach to solve Boussinesq equation is to drop 
notion of pursuing an expansion in powers of $\delta$ and instead use
weakly dispersive expression for $\phi$ or horizontal velocity in the form
of power series in $\mu^{\rm 2}$ to evaluate complete surface boundary condition. 
We refer to this procedure as $\it{ fully nonlinear}$ in the sense that all of the availble
information on velocities is used to evaluate boundary conditions. 
Conventional time dependent Boussinesq equations for surface wave
height and consequent breaking term calculation are very straight
forward and have benn calculated in case of irrotational waves.  Here
we take up fully nonlinear calculation as vorticity becomes a large
fraction of water depth in the surf zone or shoaling waves. While
developing Boussinesq equations for horizontal momentum, we retain up
to order O($\delta^{\rm 2})$ and O($\delta\mu^{ 2})$ in our fully
nonlinear calculation. Fully nonlinear Boussinesq equations for long
wave have been derived by Mei [1983] for flat bottom and by Wei et
al \cite{wei} for variable bottom surface in case of irrotational
wave.  Shen \cite{shen} addressed problems in developing generalized
three dimensional irrotational propagating wave field to include
rotational motion in general but did not describe the vorticity breaking
terms.  For horizontal propagation of waves, the three dimensional
problem can be reduced in terms of two horizontal velocity by
integrating over depth and retaining up to order $ O(\delta^{\rm 2}) $
and $ O(\delta\mu^{\rm 2})$ As horizontal velocity is governed by
momentum equation at the surface $\eta$ by,
\begin{equation}
 \cfrac{Du}{Dt} {\vert}_{\rm \eta} = (\cfrac{Dw}{Dt} {\vert}_{\rm \eta} + 1)
{\bigtriangledown} \eta 
\end{equation}

In the surf zone region of sloping beach, waves break due to high
vorticity and the breaking of wave later being converted to
turbulence. So horizontal variation of water depth ${\it h(x,y)} $
must be considered in this case.  We express surface propagation
equation in terms of average velocity description and total time
derivative of horizontal momentum can be written as,
\begin{equation}
\cfrac {D \bar u}{Dt} {\vert}_{\rm \eta} 
  =  \cfrac{\partial u}{\partial t}{\vert}_{ \eta} + u_{\rm \eta}\cdot 
(\bigtriangledown u) {\vert}_{\rm \eta} 
\end{equation}
where surface velocity is given by, 
\begin{equation}
\begin{split}
u_{\rm \eta} = \bar{u} + {\eta} s_{\rm \eta} -
\cfrac{\mu^{\rm 2}}{3}d^{\rm 2} \bigtriangledown \left (
\bigtriangledown \cdot \tilde{u} - {\eta} {\bigtriangledown} \cdot s_{\rm \eta} \right )  
+ \cfrac{\mu^{\rm 2}}{2}d \{  {\bigtriangledown} \left ( \tilde{u} -
 {\Delta} \bar{u} {\vert}_{\rm {-h}}  
 + {\eta} s_{\rm \eta} \right ) \cdot {\bigtriangledown}h \} \\
\end{split}
\end{equation}

We consider $ \bigtriangledown H_{ \eta} = \bigtriangledown \eta +
\bigtriangledown h $ for wavy bottom

\begin{equation}
\begin{split}
\cfrac{Du}{Dt}|_{\rm \eta} = \cfrac{\partial \tilde u}{\partial t}  +
+ \eta \cfrac{\partial s_{\rm \eta}}{\partial t}
+ \tilde u \cdot \bigtriangledown \tilde u 
\end{split}
\end{equation}

\begin{equation}
\begin{split}
\cfrac{Du}{Dt}|_{\rm \eta} 
=  \cfrac{\partial \tilde u}{\partial t}  +
+ {\eta} \cfrac{\partial s_{\rm \eta}}{\partial t}
+ \tilde u \cdot \bigtriangledown \tilde u 
- \cfrac{\mu^{\rm 2}}{3}d^{\rm 2} \{
{\bigtriangledown}(\bigtriangledown
\cdot \cfrac{\partial \tilde u}{\partial t} - {\eta}{\bigtriangledown}
\cdot \cfrac{\partial s_{ \eta}}{\partial t}) 
+ \tilde u \cdot \bigtriangledown 
 ( \bigtriangledown \cdot {\tilde u} 
-{\eta}{\bigtriangledown} \cdot s_{\rm \eta}) \\
+ \cfrac{\mu^{\rm 2}}{2}d \{\bigtriangledown(\cfrac{\partial \tilde u}
{\partial t} + {\eta}\cfrac{\partial s_{\rm \eta}}{\partial t} )
\cdot {\bigtriangledown} h   \\
-(\bigtriangledown \cdot \cfrac{\partial \tilde u}
{\partial t} 
- {\eta}\bigtriangledown \cdot \cfrac{\partial s_{\rm \eta}}
{\partial t}){\bigtriangledown} h  
+ \tilde u \cdot \bigtriangledown \}  + O(\mu^{ 4})
\end{split}
\end{equation}

\
This long wave momentum equation upon simplification over flat bottom
case can be compared to the one derived by Shen [2000] The
vertical velocity can be obtained similarly, \\
\begin{equation}
\cfrac {Dw}{Dt} |_{ \eta} = \cfrac{\partial w }
{\partial t} |_{\rm \eta} + 
u_{ \eta} \cdot \bigtriangledown w_{ \eta} + 
w \cfrac{\partial w}{\partial z}
|_{\rm \eta}    
\end{equation}

So, we can write the horizontal momentum equation as, 
\begin{equation}
\begin{split}
\cfrac{\partial \tilde u}{\partial t}  + \tilde u \cdot \bigtriangledown
\tilde u  + \bigtriangledown \eta
= \cfrac{\mu^{\rm 2}}{3}d^{\rm 2}   \{ \bigtriangledown(\bigtriangledown
\cdot \tilde u)\cdot \bigtriangledown 
 \tilde{u}  
+ \bigtriangledown(\bigtriangledown \cdot 
\cfrac{\partial \tilde u}{\partial t}) \\
+  (\tilde u 
\cdot \bigtriangledown(\bigtriangledown \cdot \tilde u)
\} - \mu^{\rm 2}d \{ \bigtriangledown \cdot \frac{\partial \tilde u}
{\partial t} -d^{\rm 2} \bigtriangledown(\bigtriangledown \cdot \cfrac
{\partial \tilde u}{\partial t}) + (\bigtriangledown 
\cdot u_{ \eta})^{\rm 2}  \\
- \tilde u \cdot \bigtriangledown (\bigtriangledown \cdot \tilde u) \} 
 \bigtriangledown {\eta}   
\end{split}
\end{equation}

 $ \tilde u $ is defined in previous section.  In contrast to the
 result by Shen \cite{shen}, additional contribution factor here
 arises from vorticity variation which is significant for surf zone
 wave.  Wei et al \cite{wei} also calculated breaking term for irrotational long
 wave momentum equation over a variable bottom wave. The intermediate
 depth velocity $ z_{\rm \alpha} $ is being used there proportional to
 h instead of depth average velocity used here which may not be valid
 inside the fluid.  The use of $ z_{\rm r} $ in our approach avoids
 this difficulty. Finally we try to generalize equation by solving
 vorticity from vorticity transport equation in next section.

\section{ Vorticity transport equation in breaking zone} 
Madsen and Svendsen \cite{madsen} used a cubic vertical distribution
of rotational velocity based on roller jump data which can not
considered in three dimension case as it is not guaranteed to bring
accuracy in the simulation.  So we try to solve vorticity function
from Reynold stress based equation.
\begin{equation}
\cfrac{\partial u}{\partial t} + ( u \cdot  
\bigtriangledown) u = -\cfrac{1}{\rho}{ \bigtriangledown p}
\end{equation}
Taking the curl on both sides and use vorticity function $ s =
\bigtriangledown \times u $ we get,

\begin{equation}
\cfrac{\partial s}{\partial t} -(s \cdot  
\bigtriangledown)s
+ ( u \cdot \bigtriangledown) s = {\nu}
{\bigtriangledown}^{\rm 2} s
\end{equation}

$(s \cdot \bigtriangledown) s $ is called "vorticity stretching" factor and is
 the gradient in vorticity value. This term leads to change of
rotation of material particles present in the flow to the
beach. Contribution of this term can not be incorporated from two
dimension roller jump data.
 
We generalize the equation in three dimension as
\begin{equation}
\begin{split}
\cfrac{\partial s} {\partial t}  
+ {\delta u} 
\cfrac{\partial s}{\partial x} + {\delta v} \cfrac{\partial s}{\partial y} 
+ {\delta w} \cfrac{\partial s}{\partial z} 
- {\delta s} \cfrac{\partial u}{\partial x} - {\delta s} \cfrac{\partial v}
{\partial y} -{\delta s} \cfrac{\partial w}{\partial z}  \\
= \nu \left [ {\mu}^{\rm 2}\cfrac{\partial^{\rm 2} s}{{\partial x}^{\rm 2}}
+ {\mu}^{\rm 2} \cfrac{\partial^{\rm 2} s}{{\partial y}^{\rm 2}}
+ \cfrac{\partial^{\rm 2} s}{{\partial z}^{\rm 2}} \right ]
\end{split}
\end{equation}
 
After changing the variable from (x,y,z,t) to wave following
coordinates $ (x,y,\sigma,t)$, we write the vorticity equation as
\begin{equation}
\begin{split}
\cfrac{\partial s}{\partial t} - \cfrac{\delta \sigma}{(h+\delta\eta)} 
\cfrac
{\partial \eta}{\partial t} \cfrac{\partial s}{\partial \sigma} + 
\delta u (\bigtriangledown \cdot s) - 
{\delta}s (\bigtriangledown \cdot u) 
- \cfrac{\delta }{(h+\delta \eta)} \left [ s\cfrac{\partial w}
{\partial \sigma} - w \cfrac {\partial s}
{\partial \sigma} \right ] \\
- {\delta}^{\rm 2} \frac{\sigma u }
{(h+\delta \eta)} (\bigtriangledown  \cdot
\eta)  \cfrac{\partial s}{\partial \sigma} + {\delta}^{\rm 2}
\cfrac{\sigma s}{(h+\delta\eta)}(\bigtriangledown \cdot \eta) 
\cfrac{\partial u}{\partial \sigma}  \\
= \nu \left [ {\mu^{\rm 2}} \bigtriangledown^{\rm 2}s +
 \cfrac{1}{(h+\delta \eta)}
\cfrac{\partial^{\rm 2}s}{\partial \sigma^{\rm 2}} \right ] +O(\mu^{\rm 2})
+ O(h_{\rm x}) + O(h_{\rm y})
\end{split}
\end{equation}

The boundary conditions in new coordinate system are, 
\begin{equation}
\begin{split}
s(\sigma=1,t) = s(x,y,t) ; \\
s(\sigma=0,t) = 0 ; \\
s(\sigma,t=0) = 0 
\end{split}
\end{equation}

After we redefine $ s = \Omega + \sigma {\omega}_{ s} $, which
transforms the equation,
\begin{equation}
\begin{split}
\cfrac{\partial \Omega}{\partial t} + {\sigma}\frac{\partial \omega_{ s}}{\partial t}
 - {\delta}\frac{\sigma}{(h +{\delta}{\eta})} \left [ \sigma \frac{\partial \eta}{\partial t} 
- \cfrac{\partial \eta}
{\partial t}\frac{\partial \Omega}{\partial \sigma} \right ]  \\
+ {\delta}u(\bigtriangledown \cdot \Omega) - \cfrac{\delta \sigma}
{(h+ \delta \eta)} (\bigtriangledown \cdot \eta) \cfrac{\partial \Omega}
{\partial \sigma} - \cfrac{\delta \sigma^{ 2}}{(h+\delta \eta)}
(\bigtriangledown \cdot \eta)  \\
+ \cfrac{\delta u}{(h+\delta \eta)} \cfrac{\partial \Omega}
{\partial \sigma} + \cfrac{\delta w}{(h+\delta \eta)} \omega_{ s}  \\
- \cfrac{\delta \Omega}{(h+ \delta \eta)} 
\cfrac{\partial w }{\partial \sigma} - \cfrac{\delta \sigma 
\omega_{ s}}{(h+ \delta \eta)} \cfrac{\partial w}{\partial \sigma} \\
-  \delta \Omega (\bigtriangledown \cdot u) - {\delta}{\sigma}
\omega_{ s} (\bigtriangledown \cdot u) + \cfrac{ \delta^{ 2} \sigma
\Omega}{(h + \delta \eta)}(\bigtriangledown \cdot \eta) \cfrac{\partial u}
{\partial \sigma} + \cfrac{\delta^{ 2} \sigma \Omega}{(h+\delta \eta)}
(\bigtriangledown \cdot \Omega)\cfrac{\partial u}{\partial \sigma}  \\
+ \frac{ \delta^{ 2} \sigma \omega_{ s}}{(h+\delta \eta)}
(\bigtriangledown \cdot \eta) \cfrac{\partial u}{\partial \sigma} 
\end{split}
\end{equation}  with new boundary, 

\begin{equation}
\begin{split}
\Omega(\sigma=1, t) = 0 \\
\Omega(\sigma=0, t) = 0  
\end{split} 
\end{equation}

with initial condition $ \Omega(\sigma,t=0) = 0 $. This additional
equation can be solved numerically as done by Briganti et al. \cite{briganti}
for the two dimensional case or an analytical solution can
be formulated as shown by Veeramony \& Svendsen \cite{veeramony2}.  The
analytical solution can be calculated by assuming $\Omega =
\omega^{\rm {(1)}} + {\delta}{\omega}^{\rm {(2)}} $ which gives first and
second solution as
\underline {O(1) Problem}
\begin{equation}
\cfrac{\partial \omega^{\rm  {(1)}}}{\partial t} + {\sigma} \cfrac{\partial \omega_{\rm s}}{\partial t} = 
\cfrac{\nu}{h^{\rm 2}} \cfrac{\partial^{\rm 2}\omega^{\rm {(1)}}} {\partial t^{\rm 2}}
\end{equation}

where the solution is 

\begin{equation}
 F_{\rm n}^{\rm {(1)}} = (-1)^{\rm n} \cfrac{2}{n\pi} \cfrac{\partial {\omega}_{\rm s}}{\partial t} 
\end{equation}
assuming $ - \sigma \cfrac{\partial {\omega}_{\rm s}}
{\partial t} = {\sum_{\rm n=1}}^{\infty} F_{\rm n}^{\rm 1} 
\sin (n\pi\sigma)   $  
And to solve $ \omega^{\rm {(1)}} $, assume 
$ \omega^{\rm {(1)}} = \Sigma_{\rm n} 
G_{\rm n}^{\rm {(1)}} sin (n\pi\sigma) $ 
which gives  zeroth order solution as 
\begin{equation}
G_{\rm n}^{\rm {(1)}} = (-1)^{\rm n} \cfrac{2}{n\pi} \int_{\rm 0}^{\rm t} 
\cfrac{\partial \omega_{\rm s}}{\partial \tau} e^{\rm {n^{\rm 2}}{\pi}^{\rm 2}{\kappa(\tau-t)}} d {\tau}
\end{equation}

To consider \underline { O($\delta$)Problem } 
\begin{equation}
\cfrac{\partial \omega^{ (2)}}{\partial \sigma}  - \cfrac{\nu}{h} \cfrac{\partial^{ 2} \omega^{\rm (2)}}
{\partial \sigma^{\rm 2}}  = F^{\rm (2)} 
\end{equation}

where 
\begin{equation} 
\begin{split}
F^{\rm {(2)}} = \cfrac{\sigma}{h}\cfrac{\partial \eta}{\partial t} 
\cfrac{\partial \omega^{(1)}}{\partial \sigma}
 - \cfrac{\sigma^{\rm 2}}{h} \cfrac{\partial \eta}{\partial t} 
-\cfrac{\sigma}{h} (\bigtriangledown \cdot \eta) \cfrac{\partial 
\omega^{\rm (1)}}{\partial \sigma}  - \cfrac{\sigma^{\rm 2}}
{h}(\bigtriangledown \cdot \eta)  + u( \bigtriangledown \cdot 
\omega^{\rm (1)}) \\ 
+ \cfrac{u}{h} \cfrac{\partial \omega^{\rm (1)} }{\partial \sigma} 
\end{split}
\end{equation}

 To solve above equation, assume
$ \omega^{\rm (2)} = \Sigma_{ n}^{ {(2)}} 
sin (n\pi \sigma) $ 
where solution becomes
\begin{equation}
G_{\rm n}^{\rm {(2)}} = 2\int_{ 0}^{\rm 1} 
F_{\rm n}^{\rm {(2)}}e^{ n^{ 2}\pi^{\rm 2}\kappa(\tau -t)}d {\tau}
\end{equation} with
\begin{equation}
F_{\rm  n}^{\rm {(2)}} = 2{\int}_{\rm 0}^{\rm 1} 
F^{\rm (2)}sin (n{\pi}{\sigma}) d{\sigma} 
\end{equation}
The solution for vorticity s becomes,

\begin{equation}
s = {\sigma} {\omega}_{\rm s} + \Sigma_{\rm 1} {G_{\rm n}}^{ \rm (1)}
sin (n{\pi}{\sigma}) + {\Sigma}_{\rm 1} {G_{\rm n}}^{\rm (2)} 
sin (n{\pi}{\sigma})
\end{equation}

To solve breaking term, we need value of $ \omega_{\rm s} $ for
boundary and eddy viscosity value as input data. The parametric form of vertical profile of
eddy viscosity 
is being applied to calculate vorticity function using various 
experimental data. The eddy viscosity distribution over the water column $N(z)$ is assumed such that its 
maximum value is located at the water surface except at the roller where the maximum is located
at the lower edge of the roller. The value is estimated by a mixing length hypothesis and 
$\nu_{\rm t}$ is given by, 
\begin{equation}
\nu_{\rm t}(z,x) = \nu_{\rm {0t}}h(x) \sqrt{gh(x)}N(z) 
\end{equation}
The best fit for $\omega_{\rm s}$ based on experimental data
for hydraulic jump \cite{svendsen},
\begin{equation}
\omega_{\rm s}= 15.75(1-\cfrac{x-x_{\rm t}}{l_{\rm r}})(1-e^{\rm {-40\cfrac{x-x_{\rm t}}{l_{\rm r}}}})
\end{equation}
where $x_{\rm t}$ is the position of the toe of the roller and $l_{\rm r}$ is the roller length. 
Figure 1. shows our analytical calculation time evolution of vorticity profile for $\nu_{\rm {t0}}=0.03$. 
\begin{figure}[ht]
\includegraphics[scale=0.50]{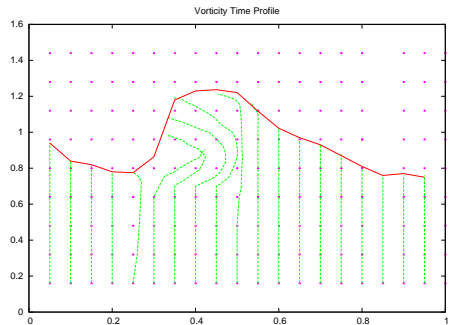}
\caption{Figure 1. shows analytical calculation for time evolution of vorticity profile}
\end{figure}
\section{Conclusion}
Finally we conclude here by developing a most generalized form 
fully nonlinear Boussinesq equations for wave propagation in surf zone
region with variable bathymetry. The vorticity distribution was calculated using
Vorticity Transport Equation(VTE). 
In the wave breaking zone, vorticity generated by the shear stress of
current is very strong, we showed in our calculation the contribution to the surface velocity due to
vorticity variation  which has significant contribution in fluid flow.
These extra terms in generalized equation complicate the numerical
technique as these terms are present in the equation in multiple form
of equations for vorticity components which has to be solved in
coupled solution technique. Veeramony \cite{veeramony1} used simplified 
formulation by taking constant eddy viscosity value but this
oversimplified case may bring inaccuracy in calculation. Briganti
et.al \cite{briganti} formulated a numerical technique scheme to solve
VTE  using generalized depth variable eddy
viscosity  $ \nu= \nu(x,y) $ in two dimension case. In three
dimensional formulation, the nonlinear terms in the vorticity
transport equation(VTE) will complicate the calculation and so proper
numerical technique have to be developed. Turbulence develops because of the instability
of vortical flow, turbulence injection can be explained from vortical function development. 
\section{list of Symbols}
$a$ -- Wave Amplitude;
$l$ -- Characteristic wave length;
${ \bf s} $ -- Vorticity function;
$g$ -- gravitational Constant;
$f$ -- bottom friction constant
\section{Acknowledgment}
Finnaly we speacially R. Dutta thanks Office of Naval Research Lab (NRL) under grant
 [GR001820] for financial support of this work.

\end{document}